\documentclass[letterpaper, conference,10pt] {IEEEtran}
\usepackage{amsmath}

\usepackage{amssymb,amsthm}
\usepackage{bm}
\usepackage{color}
\usepackage{graphicx}
\usepackage{empheq}
\usepackage[linesnumbered,ruled,lined]{algorithm2e}
\usepackage[outdir=./]{epstopdf}
\usepackage{caption}
\usepackage{lipsum}
\usepackage{cite}
\usepackage{multirow}
\usepackage{subcaption}
\usepackage{fancyhdr}
\SetKwInput{Input}{input}
\SetKwInput{Output}{output}

\usepackage{array}
\newcolumntype{L}[1]{>{\raggedright\let\newline\\\arraybackslash\hspace{0pt}}m{#1}}
\newcolumntype{C}[1]{>{\centering\let\newline\\\arraybackslash\hspace{0pt}}m{#1}}
\newcolumntype{R}[1]{>{\raggedleft\let\newline\\\arraybackslash\hspace{0pt}}m{#1}}

\theoremstyle{remark}

\IEEEoverridecommandlockouts
\def\specialpapernotice#1{\if@confmode%
	\def\@specialpapernotice{{\sublargesize\textit{#1}\vspace*{1em}}}%
	\else%
	\def\@specialpapernotice{{\\*[1.5ex]\sublargesize\textit{#1}}\vspace*{-2ex}}%
	\fi}

\begin{document}

	\title{Fully Connected Reconfigurable Intelligent Surface Aided Rate-Splitting Multiple Access for  Multi-User Multi-Antenna Transmission
   }

	\author{	\IEEEauthorblockN{Tianyu Fang$ ^{*\dagger\ddagger} $ , Yijie Mao$ ^* $, Shanpu Shen$ ^\S $, Zhencai Zhu$ ^\ddagger $, Bruno Clerckx$ ^\P $ }
		\IEEEauthorblockA{
			$ ^* $School of Information Science and Technology, ShanghaiTech University, Shanghai, China\\
			$ ^\dagger $University of Chinese Academy of Sciences, Beijing, China\\
			$ ^\ddagger $Innovation Academy for Microsatellites, Chinese Academy of Sciences, Shanghai, China\\
			$ ^\S $Department of Electronic and Computer Engineering, The Hong Kong University of Science and Technology, Hong Kong\\
			$ ^\P $Department of Electrical and Electronic Engineering,	Imperial College London, United Kingdom\\
			Email: \{fangty, maoyj\}@shanghaitech.edu.cn, sshenaa@connect.ust.hk, zczhu@hotmail.com,    b.clerckx@imperial.ac.uk}

		\thanks{
		This work was sponsored by Shanghai Sailing Program under Grant 22YF1428400.}

	\\[-3 ex]		
	}
\maketitle

\thispagestyle{empty}
\pagestyle{empty}
\begin{abstract}
Rate-splitting multiple access (RSMA) has been recognized as a promising and powerful multiple access (MA) scheme, non-orthogonal transmission framework and interference management strategy for 6G. Inspired by the appealing spectral efficiency gain achieved by RSMA over conventional MA schemes in multi-user multi-antenna transmission, in this paper we introduce RSMA to reconfigurable intelligent surface (RIS)-aided multiple-input single-out (MISO) broadcast channel (BC). To further enhance the spectral efficiency, a more generalized RIS architecture called fully connected RIS is considered. By jointly optimizing the scattering matrix of the fully connected RIS and the transmit beamformers to maximize the sum-rate, we show that the proposed fully connected RIS aided RSMA transmission scheme significantly improves the spectral efficiency compared with the conventional single connected RIS schemes and the schemes without RIS. It acts as a new benchmark for linearly precoded multi-user multi-antenna networks. 
\end{abstract}


\section{Introduction}

\label{sec:introduction}
\par  The past few years have witnessed the development of rate-splitting multiple access (RSMA) for multi-antenna networks. It has been recognized as a promising physical-layer non-orthogonal transmission strategy, a powerful interference management approach, and a candidate of the multiple access technique for the sixth generation wireless network (6G) \cite{dizdar2021rate,9390169,mao2022rate}. By splitting the user messages into common and private parts, encoding the common parts into the common streams to be decoded by multiple users, and encoding the private parts respectively into the private streams to be decoded by the corresponding users only, RSMA enables a flexible interference management capability of partially decoding the interference and partially treating the interference as noise \cite{mao2018,SE2016}. Existing works have shown that RSMA outperforms other multiple access techniques (including linearly precoded space division multiple access--SDMA, power-domain non-orthogonal multiple access--NOMA, orthogonal multiple access-OMA, and multicasting) in terms of spectral efficiency \cite{mao2018,sumrate2016,SE2016,SEandEE2019,clerckx2021noma}, energy efficiency \cite{EEmao2018,SEandEE2019}, max-min fairness \cite{clerckx2021noma,mao2020max}, and robustness towards inaccuracies of channel state information at the transmitter (CSIT) \cite{sumrate2016,mao2020beyond}.

\par Reconfigurable intelligent surfaces (RISs), consisting of a large number of reconfigurable scattering elements, is another promising technology for 6G \cite{wu2021intelligent}. RISs can smartly reconfigurable the wireless propagation environment so as to effectively enhance the spectral and energy efficiency. Most of existing RIS research relies on using a simple architecture referred to as single connected RIS, which is characterized by a diagonal matrix with constant modulus entries \cite{irs2019,FP2020,yu2021smart,wu2021intelligent}. To enhance the RIS performance, a more general architecture referred to as fully connected RIS has recently been proposed in \cite{IRSmodel2021}. The fully connected RIS is characterized by a complex symmetric unitary matrix, which achieves a better performance than the single connected RIS. 

\par The appealing performance benefits of RSMA and RIS have motivated the study on the integration of them \cite{EEwithIRS2020,MMFwithIRS2021,RSwithIRSinCloud-RAN2021,outagewithIRS2021,Bansal2021Rate}. The interplay of RSMA and RIS is first investigated in \cite{EEwithIRS2020} for multi-user multi-antenna networks, where RIS aided RSMA transmission model achieves higher energy efficiency than NOMA and orthogonal frequency division multiple access (OFDMA). RIS aided RSMA transmission is further shown to enhance the fairness among users \cite{MMFwithIRS2021}, reduce the transmit power \cite{RSwithIRSinCloud-RAN2021}, and achieve superior outage performance \cite{outagewithIRS2021,Bansal2021Rate} over conventional RIS-aided networks. However, all the above works only use the single connected RIS architecture. Therefore, we would like to further enhance the performance by leveraging the more advanced fully connected RIS architecture. To the best of our knowledge, there is no existing work that investigates the sum-rate achieved by the fully connected RIS aided RSMA transmission networks.

\par In this work, we propose a fully connected RIS aided RSMA downlink multi-antenna multi-user transmission model. The scattering matrix of RIS and the transmit beamformers of RSMA are jointly optimized in order to maximize the sum-rate. To solve the problem, we propose an alternative optimization framework where the scattering matrix of RIS and the beamformers of RSMA are iteratively optimized. Numerical results show that by synergizing RSMA and fully connected RIS, the proposed scheme significantly improves the spectral efficiency in multi-user multiple-input single-output (MU-MISO) MU-MISO networks. The proposed scheme explores a larger achievable sum-rate than the conventional single connected RIS aided schemes and the schemes without RIS. It acts as a new benchmark for linearly precoded multi-user multi-antenna networks.

\par \textit{Notations:} Vectors and matrices are denoted by bold lower and upper case letters. $ \mathbb{R}^{m \times n} $ and $ \mathbb{C}^{m\times n} $ represent the real-valued and complex-valued spaces with dimension $ m \times n  $. $|x|$ indicates the magnitude of a complex number $ x $ and $ \Re(x) $ denotes its real part. For a vector $ \mathbf{x}  $, $ \| \mathbf{x}\| $ denotes its Euclidean norm. $\mathbb{E}\{\cdot\}$ denotes the statistical expectation operator for a random variable. $(\cdot)^H$, $(\cdot)^T$ and $\mathrm{tr}(\cdot)$  respectively denote the conjugate transpose, transpose and trace operators. $ \mathrm{diag} \left( x_1,x_2,\cdots,x_n \right) $ is a diagonal matrix with  $ (x_1,x_2,\cdots,x_n) $  being its diagonal elements. $ \mathbf{I}$ is the identity matrix. $\mathcal{CN}(\mu,\sigma^2 ) $ denotes the circularly symmetric complex Gaussian (CSCG) distribution with mean $ \mu $ and variance $\sigma^2$. 
\section{System Model and Problem Formulation}
\label{sec: system model}
\par
\begin{figure}[t!]
	
	\centering
	\includegraphics[width=0.5\textwidth]{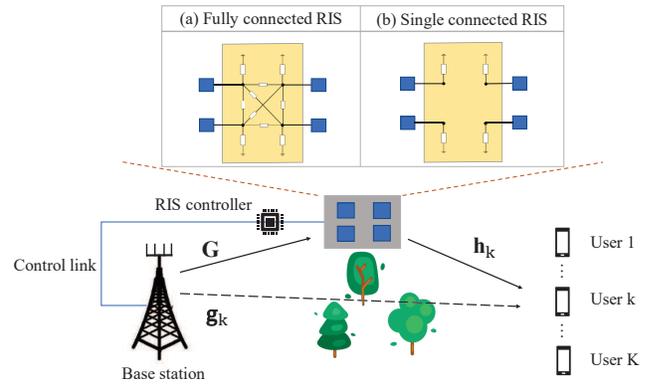}
	\caption{A multi-antenna multi-user transmission network with the assistance of a (a) fully connected RIS, (b) single connected RIS.}
	\label{fig:system}
\end{figure}

   	We consider a  MU-MISO communication network consisting of one base station (BS) equipped with $ M $ antennas, one RIS with a set of $ N $ passive reflecting elements indexed by $ \mathcal{N}=\{1,2,\ldots,N\} $, and a set of $ K $ single-antenna users indexed by $ \mathcal{K}=\{1,2,\ldots,K\} $. As illustrated in Fig.$\, \ref{fig:system} $, the BS simultaneously serves the $ K $ users with the assistance of a fully-connected RIS (as in Fig.$\, \ref{fig:system} \,$(a)). The reconfigurable impedance network of RIS is adjusted and determined by a smart controller attached to the RIS, which also acts as a gateway to exchange the information between the BS and the RIS. The channels from the BS to the users, from the RIS to the users, and from the BS to the RIS are denoted as $  \mathbf{g}_k\in \mathbb{C}^{M\times 1} $, $ \mathbf{h}_k\in\mathbb{C}^{N\times 1} $, $ k\in\mathcal{K} $ and $ \mathbf{G}\in\mathbb{C}^{N\times M} $, respectively. All channels are assumed to be invariant during one transmission block and perfect CSI is available at the BS. Although the assumption of perfect CSI is ideal, the proposed scheme explores a larger achievable sum-rate than the conventional schemes, which therefore acts as a new benchmark for multi-user multi-antenna networks as well as future study for the corresponding imperfect CSIT settings. At the BS, message $ W_k $ intends to user $ k $ is split into a common part $ W_{c,k} $ and a private part $ W_{p,k} $. The common parts of all users are combined and encoded into a common stream $ s_0 $ while the private parts are independently encoded into the private streams $ s_1,\cdots,s_K $. Denote $ \mathbf{s}=[s_0,s_1,\cdots,s_K]^T $ and $ \mathbf{W}=[\mathbf w_0,\mathbf w_1,\cdots,\mathbf w_K] \in \mathbb{C}^{M\times (K+1)} $ as the data stream vector and beamforming matrix for all streams, respectively. We assume that each stream $ s_k, $  $ k\in\mathcal{K}\cup\{0\} $ has zero mean and unit variance, i.e., $ \mathbb{E}\{\mathbf{s}\mathbf{s}^H\}=\mathbf{I} $. The transmitted signal at the BS is
\begin{equation}
	\mathbf{x}=\sum\limits_{k=0}^{K}\mathbf{w}_ks_k,
\end{equation}
and the transmit power constraint is
\begin{equation}
	\mathrm{tr}(\mathbf {WW}^H)\leq P_t,
\end{equation}
where $ P_t $ refers to the maximum transmit power of the BS.   
The signal is transmitted through the direct signal path from the BS to the users as well as the RIS-aided path. At user $ k $, the total received signal is    
\begin{equation}
	y_k=(\mathbf g_k^H+\mathbf h_k^H\mathbf \Theta\mathbf G)\sum\limits_{i=0}^{K}\mathbf w_is_i+z_k,
\end{equation} 
where $ \mathbf \Theta\in\mathbb{C}^{N\times N}  $ refers to the scattering matrix of the $ N $-port reconfigurable impedance network in the $ N $-element RIS, and $ z_k\sim \mathcal{CN}(0,\sigma_k^2) $ is the additive white Gaussian noise (AWGN).

\subsection{Fully Connected Reconfigurable Intelligent Surface}
\label{sec: IRSModel}
\par 
In this work, we focus on using a fully connected RIS \cite{IRSmodel2021} to enhance the spectral efficiency. An example of a 4-element fully connected RIS is illustrated in Fig.$ \,\ref{fig:system}\, $(a). In the reconfigurable impedance network of fully connected RIS, each port is connected with other ports through a reconfigurable reactance. Accordingly, the scattering matrix of a fully connected RIS $ \mathbf \Theta $ satisfies the constraints
\begin{subequations}
	\label{eq:scatter matrix constraints}
	\begin{align}
		&\mathbf \Theta^H\bm \Theta=\mathbf I,\\
		&\mathbf \Theta=\mathbf \Theta^T.
	\end{align}
\end{subequations}
As per \cite{IRSmodel2021,microwave2009}, constraint (\ref{eq:scatter matrix constraints}) is equivalent to
\begin{subequations}
	\label{eq:unconstrainted}
	\begin{align}
		\mathbf \Theta&=(j\mathbf X+Z_0\mathbf I)^{-1}(j\mathbf X-Z_0\mathbf I),\\
		\mathbf X&=\mathbf X^T,
	\end{align}
\end{subequations}
where $ Z_0 $ refers to the reference impedance and $ \mathbf X $ is a symmetric real matrix referring to the reactance matrix of the reconfigurable impedance network in RIS. With constraint (\ref{eq:unconstrainted}), a closed-form expression for scattering matrix $ \mathbf \Theta $ satisfying constraint (\ref{eq:scatter matrix constraints}) is obtained by introducing an unconstrained symmetrical real matrix $ \mathbf X $.

\par\textbf{\textit{Remark }1.}  \textit{When each port is disconnected with other ports in the reconfigurable impedance network, the fully connected RIS reduces to the single connected RIS \cite{kundu2021optimal} as illustrated in Fig.$ \,\ref{fig:system} \,$(b). The single connected RIS has been widely used in existing works  \cite{EEwithIRS2020,MMFwithIRS2021,outagewithIRS2021,RSwithIRSinCloud-RAN2021,Bansal2021Rate}, where the scattering matrix $ \mathbf \Theta  $ satisfies the constraint}
\begin{equation}
	\label{eq:diag scatter matrix}
	\bm \Theta=\mathrm{diag}\left( e^{j\theta_1},e^{j\theta_2},\cdots, e^{j\theta_N}        \right),
\end{equation}
\textit{where $ \theta_n\in[0,2\pi) $ denotes the phase of the scattering parameter of the $ n $-th port in reconfigurable impedance network. Accordingly, it can be also equivalently transformed to}
\begin{subequations}
	\label{eq:diag}
	\begin{align}
		\bm \Theta&=(j\mathbf X+Z_0\mathbf I)^{-1}(j\mathbf X-Z_0\mathbf I),\\
		\mathbf X&=\mathrm{diag}\left( x_1,x_2,\cdots,x_N\right), 
	\end{align}
\end{subequations}
\textit{where $ x_n \in \mathbb{R} $ is the reconfigurable reactance component connected to the $ n $-th port.} 
\par \textit{It should be noted that a $ N $-port fully connected RIS given in (\ref{eq:unconstrainted}) requires to tune $ N(N+1)/2 $ scattering parameters while a $ N $-port single connected RIS given in (\ref{eq:diag}) only requires to tune $ N $ scattering parameters. The fully connected RIS therefore brings a larger searching space for the optimal RIS design.}

\subsection{Problem Formulation}
\label{sec: PF}

\par At user sides, each user first decodes the common stream by treating all private streams as interference. Thus, the signal-to-interference-plus-noise ratio (SINR) of $ s_0 $ at user $ k $ is 
\begin{equation}
	\gamma_{0,k}=\frac{|(\mathbf g_k^H+\mathbf h_k^H\mathbf \Theta\mathbf G)\mathbf w_0|^2}{\sum\limits_{i=1}^K|(\mathbf g_k^H+\mathbf h_k^H\mathbf \Theta\mathbf G)\mathbf w_i|^2+\sigma_k^2},
\end{equation} 
and the corresponding transmission rate is $ r_{0,k}=\log_2\left(1+\gamma_{0,k}   \right) $.
To ensure common message $ s_0 $ is successfully decoded by all users, the achievable rate of $ s_0 $ should satisfy $ r_0=\min_{k\in \mathcal{K}} r_{0,k} $. After decoding the common stream $ s_0 $, each user employs SIC to remove the common stream from the received signal, and then decodes the intended private stream with the SINR
\begin{equation}\label{key}
	\gamma_k=\frac{|(\mathbf g_k^H+\mathbf h_k^H\mathbf \Theta\mathbf G)\mathbf w_k|^2}{\sum\limits_{i=1,i\neq k}^K|(\mathbf g_k^H+\mathbf h_k^H\mathbf \Theta\mathbf G)\mathbf w_i|^2+\sigma_k^2}.
\end{equation}
The rate of decoding private message is $ r_k=\log_2\left(1+\gamma_k   \right) $. User $ k $ then reconstructs its message by combining the submessages $ W_{c,k} $ and $ W_{p,k} $ respectively decoded from the common and private streams \cite{mao2018}.
	
In this work, we aim at jointly optimizing the scattering matrix of RIS $ \mathbf{\Theta} $ and the beamforming matrix $ \mathbf{W} $ of RSMA to maximize the sum-rate of the system. The sum-rate problem for the downlink fully connected RIS aided RSMA network can be formulated as:
\begin{subequations}
	\begin{align}
		(\mathcal{P}_1)\,\,	\max\limits_{\bm \Theta,\mathbf W}\,\, &\sum\limits_{k=0}^K r_k\\
	\label{eq:power constraint}	\text{s.t.}\,\,& \mathrm{tr}(\mathbf {WW}^H)\leq P_{t},\\
	\label{eq:scatter constraint 1}	&\bm \Theta^H\bm \Theta=\bm I,\\
	\label{eq:scatter constraint 2}	&\bm \Theta=\bm \Theta^T.
	\end{align}
\end{subequations}
Constraint (\ref{eq:power constraint}) is the transmit power constraint at the BS. (\ref{eq:scatter constraint 1}) and (\ref{eq:scatter constraint 2}) show that the reconfigurable impedance network in fully connected RIS is a lossless and reciprocal circuit network. 

When constraints (\ref{eq:scatter constraint 1}) and (\ref{eq:scatter constraint 2}) for scattering matrix $ \mathbf \Theta $ are replaced by constraint (\ref{eq:diag scatter matrix}), $ \mathcal{P}_1 $ reduces to the sum-rate problem of the single connected RIS aided RSMA \cite{WSRwithIRS2021}. When the power allocated to $ \mathbf w_0 $ is fixed to zero, the problem reduces to the sum-rate problem for the conventional single connected RIS aided SDMA \cite{FP2020}.

\section{Alternative Optimization Framework}
\label{sec: alternative optimization}
Problem $ \mathcal{P}_1 $ is a joint beamforming matrix and RIS scattering matrix optimization problem. It is non-convex and the beamformers are coupled with the scattering matrix in multiple fractional SINR expressions. Following existing works \cite{irs2019,FP2020,EEwithIRS2020,MMFwithIRS2021,outagewithIRS2021,RSwithIRSinCloud-RAN2021,Bansal2021Rate}, we propose an alternative optimization (AO) framework to solve $ \mathcal{P}_1 $. Specifically, the problem is first decomposed into the subproblems of beamforming design and scattering matrix design. The former is solved by a weighted minimum mean square error (WMMSE)-based approach while the latter is solved by the quasi-Newton algorithm. The two subproblems are solved iteratively until convergence. In the following subsections, the proposed optimization algorithm is delineated.  

\subsection{Beamforming Optimization}
\par  With a given scattering matrix $ \bm\Theta $, the channel responses from the RIS to the users are fixed. To ease notations, we denote the effective channel from the BS and the RIS to user $ k $ as 
\begin{equation}
	\label{eq:fixch}
	\widetilde{\mathbf g}_k^H=\mathbf g_k^H+\mathbf h_k^H\mathbf \Theta \mathbf G.
\end{equation}
And $ \mathcal{P}_1 $ reduces to 
\begin{subequations}
	\begin{align}
		(\mathcal{P}_2)\,\,	\max\limits_{\mathbf W}\,\, &\sum\limits_{k=0}^K r_k\\
		\text{s.t.} \, &  \mathrm{tr}(\mathbf {WW}^H)\leq P_{t},
	\end{align}
\end{subequations}
which can be solved by the WMMSE algorithm \cite{sumrate2016} as briefly introduced below.

\par Denote the equalizers to estimate $ s_0 $ and $ s_k $ as $ e_{0,k} $ and $ e_k $, respectively. $ \hat{s}_{0,k}=e_{0,k}y_k $ is the estimate of $ s_0 $, and $ \hat{s}_k=e_k(y_k-\widetilde{\mathbf g}_k^H\mathbf w_0 \hat{s}_{0,k}) $ is the estimate of $ s_k $ at user $ k $. The mean square errors (MSEs) of decoding $ s_0 $ and $ s_k $ are calculated as 
\begin{equation}
	\label{eq:MSEs}
	\begin{split}
		\varepsilon_{0,k} &\triangleq \mathbb{E}\{ |\hat{s}_{0,k}-s_{0,k}|^2  \} =|e_{0,k}|^2 T_{0,k}-2\Re \{e_{0,k}\widetilde{\mathbf g}_k^H\mathbf w_0\}+1,\\
		\varepsilon_{k} &\triangleq \mathbb{E}\{ |\hat{s}_{k}-s_{k}|^2  \} =|e_{k}|^2 T_{k}-2\Re \{e_{k}\widetilde{\mathbf g}_k^H\mathbf w_k\}+1,
	\end{split}		
\end{equation}
where $ 	T_{0,k}=\textstyle\sum_{i=0}^K|\widetilde{\mathbf g}_k^H\mathbf w_i|^2+\sigma_k^2$, and $
		T_k= \textstyle\sum_{i=1}^K|\widetilde{\mathbf g}_k^H\mathbf w_i|^2+\sigma_k^2 $ are the average power of the received signal and the signal after removing the common stream, respectively. By setting $  \partial \varepsilon_{0,k}/\partial e_{0,k}$ and $ \partial \varepsilon_{k} /\partial e_k$ to zero respectively, the minimum MSE (MMSE) equalizers are given by
\begin{equation}
	\label{eq:emmse}
e_{0,k}^{\text{MMSE}}=\mathbf w_0^H \widetilde{\mathbf g}_k T_{0,k}^{-1}, \,\, e_k^{\text{MMSE}}=\mathbf w_k^H \widetilde{\mathbf g}_k T_k^{-1}.
\end{equation}
Substituting (\ref{eq:emmse}) into (\ref{eq:MSEs}), the MMSEs are given by $ \varepsilon_{0,k}^{\text{MMSE}}=(T_{0,k}-|\widetilde{\mathbf g}_k^H\mathbf w_0|^2)T_{0,k}^{-1},\,\, \varepsilon_{k}^{\text{MMSE}}=(T_k-|\widetilde{\mathbf g}_k^H\mathbf w_k|^2)T_k^{-1}.$ Then, the SINRs coresponding to $ s_0 $ and $ s_k $ can be transformed to $ \gamma_{0,k}=1/\varepsilon_{0,k}^{\text{MMSE}}-1,\,\gamma_k=1/\varepsilon_k^{\text{MMSE}}-1. $ The rates of common and private streams become $  r_{0,k}=-\log_2(\varepsilon_{0,k}^{\text{MMSE}})$ and $ \,\, r_k=-\log_2(\varepsilon_{k}^{\text{MMSE}}). $ However, the logarithmic rate-MMSE relationships above cannot be used directly for the sum-rate problem. To tackle the issue, the augmented MMSEs are introduced as follows
\begin{equation}
	\label{eq:WMMSEs}
	\xi_{0,k}\triangleq \lambda_{0,k}\varepsilon_{0,k}-\log_2(\lambda_{0,k}),\,\, \xi_k\triangleq \lambda_k\varepsilon_k-\log (\lambda_k),
\end{equation} 
 where $ \lambda_{0,k} $ and $ \lambda_k $ are auxiliary variables (also known as weights) for the rate-WMMSE relationships of $ r_{0,k}  $ and $ r_k $, respectively. By calculating$  	\frac{\partial \xi_{0,k}}{\partial \lambda_{0,k}}=0,\, \frac{\partial \xi_k}{\partial \lambda_{k}}=0 $, we  obtain the optimum weights given by
 \begin{equation}
 	\label{eq:weight}
 	\lambda_{0,k}^{\text{MMSE}}=(\varepsilon_{0,k}^{\text{MMSE}})^{-1},\,\, \lambda_{k}^{\text{MMSE}}=(\varepsilon_{k}^{\text{MMSE}})^{-1}.
 \end{equation}
Substituting (\ref{eq:MSEs}) and (\ref{eq:weight}) into (\ref{eq:WMMSEs}), the rate-WMMSE relationships are established as $ \xi_{0,k}^{\text{MMSE}}=1-r_{0,k},\,\, \xi_k^{\text{MMSE}}=1-r_k. $ With the rate-WMMSE relationships above, $ \mathcal{P}_2 $ is equivalently transformed into the WMMSE problem
\begin{subequations}
	\label{P3}
	\begin{align}
		(\mathcal{P}_3)\, \min\limits_{\mathbf W, \bm \lambda, \mathbf e} \, &\sum\limits_{k=0}^K  \xi_{k}\\
		\text{s.t.} \, & \mathrm{tr}(\mathbf {WW}^H)\leq P_{t},
	\end{align}
\end{subequations} 
where  $ \bm\lambda=[\lambda_{0,1},\cdots,\lambda_{0,K},\lambda_1,\cdots,\lambda_K]^T $ is the weight vector and $ \mathbf e=[e_{0,1},\cdots,e_{0,K},e_1,\cdots,e_K]^T $ is the equalizer vector. $ \xi_0=\max_{k\in\mathcal{K}} \xi_{0,k} $. $ \mathcal{P}_3 $ is still non-convex, an AO framework is applied to decompose it into three convex subproblems. For each block, one of $ \mathbf W ,\bm \lambda,\bm e$ is optimized by fixing the other two blocks. Algorithm 1 specifies the procedure of the WMMSE method to optimize the beamforming vectors. Readers are referred to \cite{WMMSE2011} for the details of the convergence proof for Algorithm 1.
 \setlength{\textfloatsep}{7pt}	
\begin{algorithm}[t!]
	\label{alg:AO for RSMA}
	\caption{WMMSE algorithm for beamforming design}	
	 \textbf{Initialize:}  $t \leftarrow 0$, $ \epsilon $, $ \mathbf W^{[t]} $, $ \mathrm{SR}^{[t]} $\;
	 \Repeat{$ |\mathrm{SR}^{[t]}- \mathrm{SR}^{[t-1]}|<\epsilon  $} {
	 t $\leftarrow$ t+1\;
	 update $ \mathbf e^{[t]},\mathbf \lambda^{[t]}$ by (\ref{eq:emmse}), (\ref{eq:weight})\;
	 update $ \mathbf W^{[t]} $ by solving problem $ \mathcal{P}_3 $ using $ \mathbf e^{[t]},\mathbf \lambda^{[t]}$ \;
 
}
	  \label{alg:WMMSE}
\end{algorithm}

\subsection{Scattering Matrix Optimization}
\label{sec: algorithm}
\par Similarly, with a given beamforming design $ \mathbf{W} $, problem $ \mathcal{P}_1 $ is simplified as
    \begin{subequations}
	\begin{align}
		(\mathcal{P}_4)\,\,	\max\limits_{\bm \Theta}\,\, &\sum\limits_{k=0}^K r_k\\
		\text{s.t.}\,\,&\bm \Theta^H\bm \Theta=\bm I,\\
		&\bm \Theta=\bm \Theta^T.	
	\end{align}
\end{subequations} 
However, it is challenging to transform problem $ \mathcal{P}_4 $ into a convex problem due to the non-convex matrix equality constraints. Hence, we apply equality (\ref{eq:unconstrainted}) to equivalently reformulate $ \mathcal{P}_4 $ as
\begin{subequations}
\begin{align}
\label{eq:object}		
(\mathcal{P}_5)	\,	\max\limits_{\bm X
	}\, &\sum\limits_{k=0}^K r_k\\
\label{eq:scatter} \text{s.t.}\,\, &\mathbf \Theta=(j\mathbf X+Z_0\mathbf I)^{-1}(j\mathbf X-Z_0\mathbf I),\\
\label{eq:symetric}&\mathbf X=\mathbf X^T.
\end{align}
\end{subequations}
 Substituting (\ref{eq:scatter}) into (\ref{eq:object}) , and removing (\ref{eq:symetric}) (by defining a symmetric matrix variable), problem $\mathcal{P}_5$ becomes an unconstrained optimization problem. Moreover, the matrix variable $\mathbf X $ is a real symmetric matrix in which $ N(N+1)/2 $ variables are adjustable. Such unconstrained optimization problem can be directly solved by the quasi-Newton method \cite{IRSmodel2021}.

\subsection{Alternative Optimization Algorithm}
 \setlength{\textfloatsep}{7pt}	
\begin{algorithm}[t!]
	\label{alg:Iterative optimization}
	\caption{Alternative Optimization to solve ($ \mathcal{P}_1 $)}	
	\textbf{Initilize:} $n \leftarrow 0$, $ \epsilon $, $ \mathbf W^{[n]}$, $\mathbf X^{[n]}$, $ \mathrm{SR}^{[n]} $\;
	\Repeat{$ |\mathrm{SR}^{[n]}-\mathrm{SR}^{[n-1]}|<\epsilon $}{ $n\leftarrow n+1$ \;
		calculate $ \mathbf \Theta^{[n-1]} $ by (\ref{eq:scatter})\;
		Given $ \mathbf \Theta^{[n-1]} $, update $ \mathbf W^{[n]} $ by solving $ \mathcal{P}_3 $ with Algorithm 1\;
		Given $ \mathbf W^{[n]}$, update $ \mathbf X^{[n]} $ by solving $ \mathcal{P}_5 $ with the quasi-Newton method using $ \mathbf X^{[n-1]} $ as the initial point\;
	}
	
\end{algorithm}
\par The proposed AO algorithm to jointly maximize the scattering matrix and the beamforming matrix is specified in Algorithm 2. Starting with a feasible beamforming matrix $ \mathbf W^{[0]} $ and a symmetric reactance matrix $ \mathbf X^{[0]} $, in $ n $-th iteration, we first calculate the scattering matrix $ \mathbf \Theta^{[n-1]} $ with reactance matrix $ \mathbf X^{[n-1]} $ from the last iteration. Next, with a fixed scattering matrix $ \mathbf \Theta^{[n-1]} $, the beamforming matrix $ \mathbf W^{[n]} $ is updated by Algorithm 1. For a given $ \mathbf W^{[n]} $, the reactance matrix $ \mathbf X^{[n]} $ is then updated based on the quasi-Newton method using $ \mathbf X^{[n-1]} $ as the initial point. The sum-rate $ \mathrm{SR}^{[n]} $ is then calculated based on the updated $ \mathbf W^{[n]} $ and $ \mathbf X^{[n]}  $. The process is repeated until convergence. 

\par \textit{Convergence Analysis}: In each iteration $ [n] $, the solution of $ \mathcal{P}_1 $ is also a feasible solution of $ \mathcal{P}_1 $ for the next iteration. Hence, the sum-rate $ \mathrm{SR}^{[n+1]} $ is larger than or equal to $ \mathrm{SR}^{[n]} $. Moreover, the non-decreasing sequences generated by Algorithm 2 is bounded above by the transmit power constraint. Therefore, the proposed AO algorithm is guaranteed to converge with a given tolerance $ \epsilon $.

\section{Numerical Results}
\label{sec: numerical results}
\par In this section, we evaluate the performance of the proposed system model and the proposed algorithm. The following six schemes are compared:
\begin{itemize}
	\item \textbf{Fully RIS RSMA}: This is the scheme proposed in Section \ref{sec: system model}.
	\item \textbf{Fully RIS SDMA}: This is a special case of the proposed scheme when the power allocated to the common stream is fixed to zero.
	\item \textbf{Single RIS RSMA}: This is the single connected RIS aided RSMA scheme, as studied in \cite{WSRwithIRS2021}.
	\item \textbf{Single RIS SDMA}: This is the conventional single connected RIS aided SDMA scheme, as studied in \cite{FP2020}.
	\item\textbf{no RIS RSMA}: This is the conventional RSMA scheme without using RIS, as studied in \cite{mao2018,sumrate2016,SE2016,SEandEE2019,clerckx2021noma}.
	\item \textbf{no RIS SDMA}: This is the conventional multi-user linearly precoded SDMA scheme without using RIS, as studied in \cite{mao2018,christensen2008weighted}.
\end{itemize}

The sum-rate maximization problems of fully/single RIS RSMA and fully/single RIS SDMA are solved by Algorithm 2 while the corresponding problems of no RIS RSMA and no RIS SDMA are solved by WMMSE directly. Problem $ \mathcal{P}_3 $ is solved by the CVX toolbox \cite{grant2008cvx} with the interior-point method, and problem $ \mathcal{P}_5 $ is solved by the optimization toolbox in Matlab with the quasi-Newton method. 

\begin{figure}[t!]
 	
 	\centering
 	\includegraphics[width=0.4\textwidth]{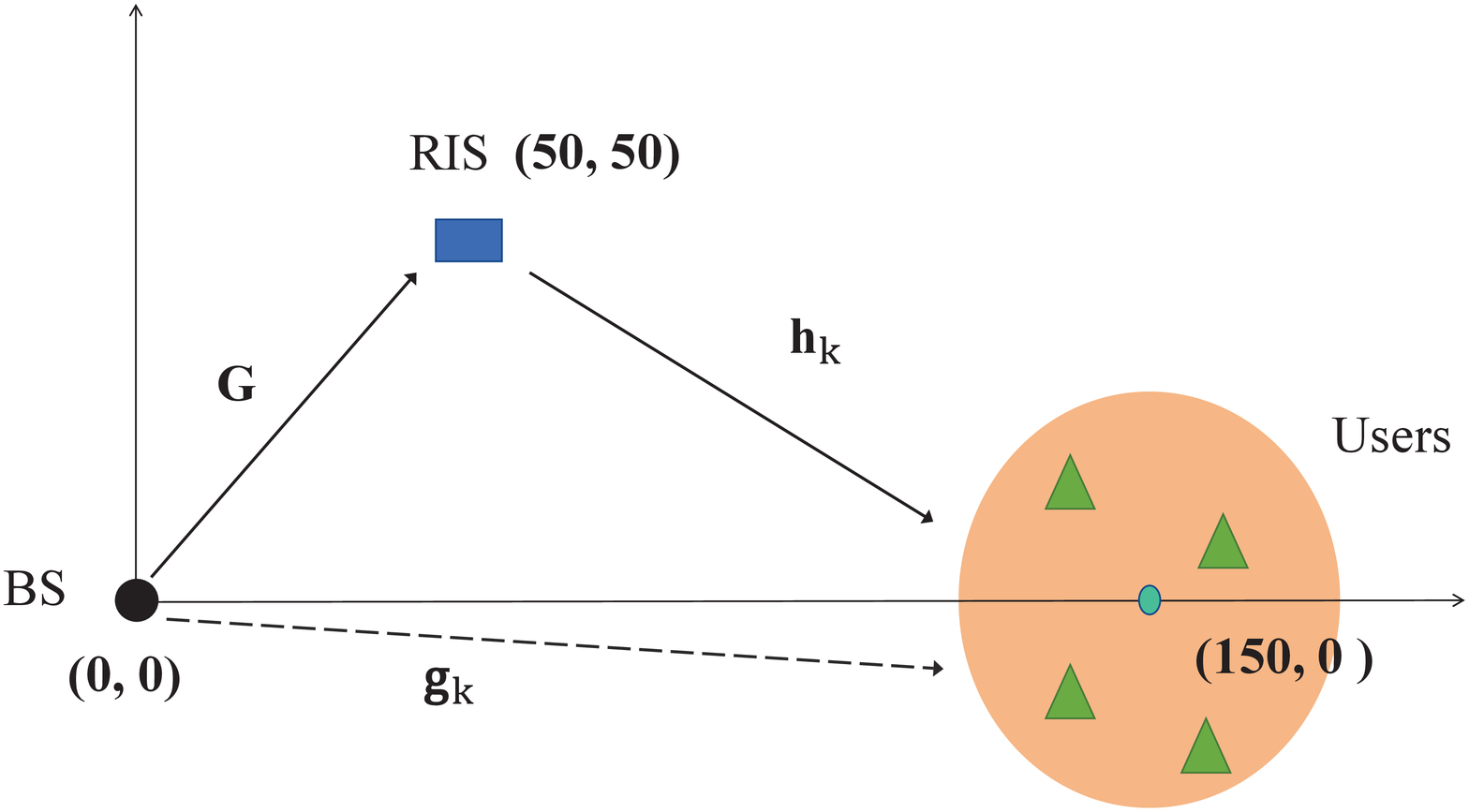}
 	\caption{The Simulated RIS-aided $ K $-user MISO transmission scenario.}
 	\label{fig:simulation}
 \end{figure}

The setting of the simulation follows \cite{irs2019 }, which is a planar RIS-aided network as shown in Fig.$\, \ref{fig:simulation} $. The BS and RIS are located at $ (0,0) $ and $ (50,50) $, respectively. In addition, there are $ K=4 $ users randomly generated in a circle centered at $ (150,0) $ meters with a diameter of 20 meters. The path loss of the channels are modeled as $ P(d)=L_0d^{-\alpha} $, where $ L_0=-30 $ dB is the reference path loss at $ d=1 $ m, $ d $ refers to the link distance, and $ \alpha $ denotes the path loss exponent. Assuming that the location of RIS is chosen carefully, we set the path loss exponent of BS to users, BS to RIS, and RIS to users are 3.5, 2, 2.2, respectively \cite{irs2019}. For simplicity, the small-scale fading of all channels are modeled as Rayleigh fading. Hence, the channels are given as $ {\mathbf g}_k\sim\mathcal{CN}(0,P(d_k^g)\mathbf I)), \mathbf{h}_k\sim  \mathcal{CN}(0,P(d_k^h)\mathbf I),$ and $ \mathbf G \sim  \mathcal{CN}(0,P(d^G)\mathbf I)$, where $ d_k^g, d_k^h $ and $ d^G $  respectively denote the distance between the BS and user $ k $, the distance between the RIS and user $ k $, and the distance between the BS and RIS. Besides, the reference impedance of RIS is $ Z_0=50 $ $ \Omega $, the convergence tolerance is $ \epsilon=10^{-3} $, and the noise at user $ k $ is  $\sigma_k^2=1  $. The transmit singal-to-noise ratio (SNR) SNR$\triangleq P_t/\sigma_k^2  $ is therefore equal to the transmit power numerically. All simulation results are averaged over 100 random channel  realizations.

 \begin{figure}[t!]
 	
 	\centering
 	\includegraphics[width=0.44\textwidth]{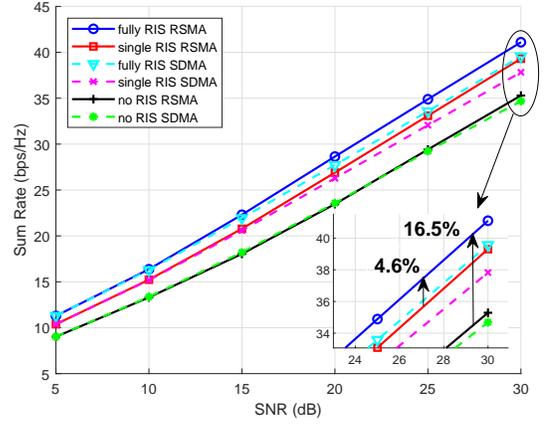}
 	\caption{Sum rate versus the transmit power, when $ M=4 $, $ K=4 $, and $ N=32 $. }
 	\label{fig:SR_vs_power}
 \end{figure}

Fig.$\, \ref{fig:SR_vs_power} $ shows the sum-rate of different strategies versus the transmit power when $ M=4 $, $ K=4 $, and $ N=32 $. It shows that the proposed fully connected RIS aided RSMA scheme outperforms all other baseline schemes. The relative sum-rate gain of fully RIS RSMA over single RIS RSMA and no RIS RSMA are at least $4.6 \%  $ and $ 16.5\% $ respectively when SNR is $ 30 $ dB. By using the fully connected RIS aided RSMA model, the sum-rate of the multi-user multi-antenna network increases significantly. Moreover, the single connected RIS aided RSMA achieves approximately the same sum-rate as the fully connected RIS aided SDMA in the high SNR regime.

\begin{figure}[t!]
	
	\centering
	\includegraphics[width=0.44\textwidth]{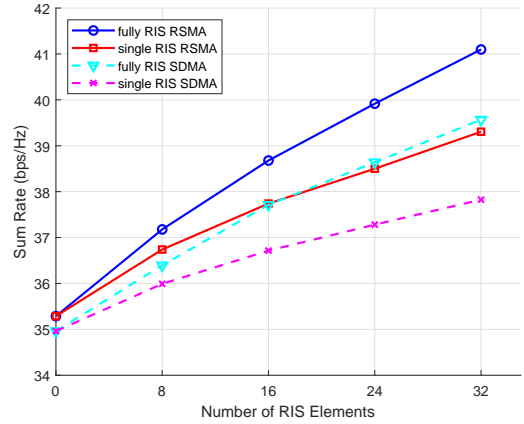}
	\caption{Sum rate versus the number of RIS Elements, when $ M=4 $, $ K=4 $, and SNR is $ 30 $ dB. }
	\label{fig:SR_vs_Nr}
\end{figure}

Fig.$\, \ref{fig:SR_vs_Nr}  $ shows the impact of the number of passive reflecting elements at the RIS (i.e., $ N $) to the sum-rate of different strategies when $ M=4 $, $ K=4 $, and SNR is $ 30 $ dB. For both fully RIS RSMA and SDMA schemes, the sum-rate increases faster than the corresponding single RIS RSMA and SDMA as $ N $ increases. Particularly, the single RIS RSMA achieves a higher sum-rate than the fully RIS SDMA when the number of elements is less than $ 16 $. In this regime, the gain obtained by RSMA scheme is more significant than the gain obtained by fully connected RIS.

\begin{figure}[t!]
	
	\centering
	\includegraphics[width=0.44\textwidth]{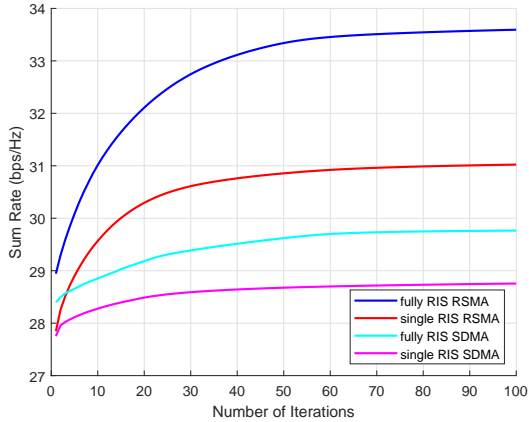}
	\caption{Convergence of the algorithms in one channel realization.}
	\label{fig:convergence}
\end{figure}
\par Fig.$\,\ref{fig:convergence} $ illustrates the convergence of Algorithm 2 for the fully RIS RSMA scheme and other baseline schemes (no RIS schemes are not included since they only adopt Algorithm 1) when $ M=4 $, $ K=4 $, $ N=32 $ and SNR is $ 25 $ dB. It can be observed that single connected RIS aided schemes converge faster than the fully connected RIS aided schemes due to the smaller number of variables in the RIS scattering matrix. In general, the algorithm can converge with 100 iterations.

\section{Conclusion}
\label{sec: conclusion}
\par In this work, we propose a fully connected RIS aided RSMA downlink transmission network. The beamforming vectors at the BS and the scattering matrix of the fully connected RIS are jointly designed to maximize the sum-rate of the network. To solve this problem, we propose an effective algorithm that alternatively optimizes the beamforming and scattering matrices. Simulation results show the outstanding spectral efficiency of the proposed  fully connected RIS aided RSMA scheme over the existing transmission schemes. It acts as a new benchmark for linearly precoded multi-user multi-antenna networks. Moreover, we show that the single connected RIS aided RSMA can achieve approximately the same spectral efficiency as the fully connected RIS aided SDMA in the high SNR regime. Therefore, we conclude that by marrying RSMA and RIS, the spectral efficiency can be enhanced significantly.

\bibliographystyle{IEEEtran}
\bibliography{reference}
\end{document}